\documentclass[12pt]{amsart}
\usepackage{amssymb,amsmath}
\usepackage{colordvi}
\usepackage[pdftex]{graphicx}
\usepackage{amsthm}
\begin{document}
\title{Comment on the Article ``Distilling   Free-Form \\ Natural  Laws from Experimental Data"}

\author{Christopher Hillar}%
\address{Mathematical Research Sciences Institute, 17 Gauss Way, Berkeley, CA 94720} 
\email{chillar@msri.org}

\author{Friedrich T. Sommer}%
\address{Redwood Center for Theoretical Neuroscience, Berkeley, CA 94720} 
\email{fsommer@berkeley.edu}

  %\date{}%
%\dedicatory{}%
%\commby{}%
% ----------------------------------------------------------------
%\begin{abstract}
  
%\end{abstract}
 \maketitle
% ----------------------------------------------------------------

\section{Summary}

A paper by Schmidt and Lipson \cite{SL,SL2} introduced the idea that ``free-form natural laws" can be learned from experimental measurements in a physical system using symbolic regression algorithms.  The most important component of that work is a fitness function involving the pair-wise derivatives of time-series measurements of system state-space data, which is purported to contain no assumptions about physical laws.  After a thorough examination of their paper \cite{SL} and supplemental materials \cite{SL2}, we submitted a technical comment to Science, which was eventually rejected.  We state the summary of the findings from our investigation of \cite{SL,SL2} here:

\begin{itemize}
\item The paper makes nonstandard use of mathematical terms and symbols such as ``dependent" and ``independent" variables, ``symbolic derivative", ``differential relationships", ``law equation", ``$\frac{\delta f}{\delta x}$", ``$\frac{\delta x}{\delta y}$", and lacks clear definitions for  concepts.
\item No theoretical justification is provided for their methods.
\item The proposed fitness function \cite[Equation S8]{SL2} is flat for general systems.
\item An alteration of their fitness function for higher order systems is able to find Hamiltonians and special classes of Lagrangians, but not general Lagrangians.
\item Previous related work is not cited.  Symbolic equation finding for time-series data appeared in \cite{JC1},  while  \cite{JC2} addressed fundamental issues with the approach.
\item A direct incorporation of Hamilton's equations into a fitness function finds the (unique) Hamiltonian of a system.  One also finds Lagrangians by incorporating the Euler-Lagrange equations into such a function.
\end{itemize}

If the fitness function for systems with more than two variables \cite[Equation S8]{SL2} is reinterpreted (as discussed in Section \ref{howresults}), then the paradigm can discover (non-canonical) compositions of Hamiltonians with differentiable functions and special classes of Lagrangians.  This follows from the specific form of the fitness function:  it is either encoding a consequence of Hamilton's equations of motion, Eqn. (\ref{pluseq}), or Newton's $2$nd law for a force arising from a potential, Eqn. (\ref{newton2ndlaw}).  In particular, the fitness directly incorporates laws of physics.  Thus, a major claim that ``[w]ithout any prior knowledge about physics $\ldots$ the algorithm discovered Hamiltonians, Lagrangians and other laws"  \cite{SL} appears to be false.  

The organization of this document is as follows.  In Section \ref{invalidfitness}, we prove mathematically that the general fitness function of the authors is inadequate.  In Section \ref{howresults}, we explain how a different fitness function than the one described in \cite[Equation S8]{SL2} might have been used by Schmidt and Lipson to obtain their results.  We also argue  how physical laws are encoded in this measure.  Finally, in Section \ref{HSfitness}, we offer another approach to the fitness which gives the unique Hamiltonian of a system as well as general Lagrangians.

\section{A flat fitness function}\label{invalidfitness}
The authors of  \cite{SL} search for ``conservation law equations'' between measured variables in a physical system by performing symbolic regression over special function classes.  The basic ideas and building blocks are represented in \cite[Figure 2]{SL}.  Given a possible conservation function $f$, its fitness with respect to the data is calculated; functions with high fitness are then mated and mutated according to standard (genetic) symbolic regression routines.  The authors' fitness measure for higher order systems \cite[Equation S8]{SL2} (as extracted from a careful reading of \cite{SL2}), however, is provably inadequate.  We first go through the mathematical details that support this assertion.  We then show how Schmidt and Lipson had carried out this argument explicitly for the Hamiltonian $f$ of a double pendulum system \cite[Section S3]{SL2}.  The authors intended for this calculation to verify that, in this case, their fitness measure correctly identified the conservation law.  As we now show, however, any function $f$ would have produced this same fitness.

Given a function $f(q_1,\ldots,q_d;\dot{q}_1,\ldots,\dot{q}_d)$ in $d > 1$ generalized coordinates, a system trajectory \[\Gamma =  \{(q_1(t),\ldots,q_d(t),\dot{q}_1(t),\ldots,\dot{q}_d(t)): t \in [0,a]\},\] and a pair of system variables $x,y \in \{q_1,\ldots,q_d,\dot{q}_1,\ldots,\dot{q}_d\}$, Schmidt and Lipson define the following three quantities (see \cite[Figure 2]{SL}, \cite[Equation S4]{SL2}, \cite[Section S2]{SL2}, and \cite[Section S3]{SL2}):

\begin{equation}\label{theirsympartderiv}
\frac{\delta f}{\delta y} := \frac{\partial f}{\partial y} + \frac{\partial f}{\partial x} \frac{d x}{d y}, \ \  \frac{\delta f}{\delta x} := \frac{\partial f}{\partial x} + \frac{\partial f}{\partial y} \frac{d y}{d x}; \ \ \frac{\delta x}{\delta y}\big{|}_{pairing} := \frac{\delta f}{\delta y}/\frac{\delta f}{\delta x}.
\end{equation}

Here, the quantity $\frac{\partial f}{\partial x}$ is standard notation for the partial derivative of the function $f$ with respect to the variable $x$ and the quantity $\frac{dx}{dy}$ (resp. $\frac{dy}{dx}$) is the rate of change of $x$ with respect to $y$ along the trajectory $\Gamma$:
\[ \frac{d x}{d y} := \frac{\dot{x}(t)}{\dot{y}(t)} = \frac{d x(t)}{d t} / \frac{d y(t)}{d t}.\]

The pair $\{x,y\}$ is called a ``variable pairing", and the ``measure of predictive ability" of a potential conservation law $f$ is given by \cite[p. 5]{SL2}:
\begin{equation}\label{SLfitness}
\min_{pairing} \left\{-\frac{1}{N} \sum_{k = 1}^N \log \left(1 + \left | \frac{d x_k}{d y_k} - \frac{\delta x_k}{\delta y_k}\big{|}_{pairing} \right| \right) \right\}. 
\end{equation}
The symbol $\frac{d x_k}{d y_k}$ (similarly for $\frac{\delta x_k}{\delta y_k}$) is evaluation of $\frac{d x}{d y}$ at discretized time step $k$ of $\Gamma$.

We claim that for \textit{any} pairing $\{x,y\}$ and \textit{any} function $f(q_1,\ldots,q_d;\dot{q}_1,\ldots,\dot{q}_d)$, the quantity $\frac{\delta x}{\delta y}|_{pairing}$ is identically equal to $\frac{dx}{dy}$ along all points of the trajectory.   This implies that the fitness function (\ref{SLfitness}) evaluates to zero for every function $f$.  
The following is the straightforward proof.  For each pairing $\{x,y\}$, we have: 
\begin{equation}\label{derivequality}
\begin{split}
 \frac{\delta x}{\delta y}\big{|}_{pairing} = \ &  \frac{\frac{\partial f}{\partial y} + \frac{\partial f}{\partial x} \frac{d x}{d y} }{\frac{\partial f}{\partial x} + \frac{\partial f}{\partial y} \frac{d y}{d x}}  = \frac{d x}{d y} \cdot \frac{ \frac{\partial f}{\partial y} \left(\frac{d x}{d y}\right)^{-1}+ \frac{\partial f}{\partial x}  }{\frac{\partial f}{\partial x} + \frac{\partial f}{\partial y} \frac{d y}{d x} } = \frac{d x}{d y} \cdot \frac{ \frac{\partial f}{\partial y} \frac{d y}{d x}+ \frac{\partial f}{\partial x}  }{\frac{\partial f}{\partial x} + \frac{\partial f}{\partial y} \frac{d y}{d x} } = \frac{d x}{d y}\cdot 1 =  \frac{d x}{d y}.  \\
\end{split}
\end{equation}

In Section S3 of \cite{SL2}, the authors provide ``an example calculation of a partial derivative pair for a double pendulum Hamiltonian" $f(\theta_1,\theta_2;\omega_1,\omega_2)$: 
\begin{equation}\label{doublehamfit}
``f = \omega_1^2+ \omega_2^2 + \omega_1 \omega_2 \cos(\theta_1 - \theta_2) - \cos \theta_1  - \cos \theta_2 ."
\end{equation}
 They compute $\frac{\delta f}{\delta \theta_1}$ and $\frac{\delta f}{\delta \theta_2}$ for the variable pairing $\{\theta_1,\theta_2 \}$, writing:  
\begin{equation}\label{theirderivex}
\begin{split}
``\frac{\delta f}{\delta \theta_1} = \ & -\omega_1 \omega_2 \sin(\theta_1 - \theta_2)\cdot \left( 1 - \frac{\Delta \theta_2}{\Delta \theta_1} \right) + \sin \theta_1 +  \frac{\Delta \theta_2}{\Delta \theta_1}  \sin \theta_2 \\
\frac{\delta f}{\delta \theta_2} = \ &  -\omega_1 \omega_2 \sin(\theta_1 - \theta_2)\cdot \left( \frac{\Delta \theta_1}{\Delta \theta_2} -1  \right) +  \frac{\Delta \theta_1}{\Delta \theta_2}  \sin \theta_1+\sin \theta_2".
\end{split}
\end{equation}
Equation (\ref{theirderivex}) is precisely definition (\ref{theirsympartderiv}) with $x$ and $y$ being $\theta_1$ and $\theta_2$, respectively.  The authors then go on to calculate that the ratio $\frac{\delta f}{\delta \theta_2} / \frac{\delta f}{\delta \theta_1}$ upon simplification is $\frac{\Delta \theta_1}{\Delta \theta_2}$.  They claim this shows that the ``partial derivative ratio resolves numerically to our estimated partial derivative pair from the experimental data, relating Eqns. (S1) and (S2)."  As we proved here in (\ref{derivequality}), every function $f$ gives an equality between ``Eqns. (S1) and (S2)" in this way.  Moreover, this is the case for each variable pairing, implying a flat fitness using the fitness function in \cite{SL,SL2} for any $f$.

\section{How Schmidt and Lipson conceivably arrived at their results}\label{howresults}

We believe that Schmidt and Lipson are choosing a fitness measure $M(f)$ with the following property: it is very large for (potential law) functions $f = f(q_1,\ldots,q_d;\dot{q}_1,\ldots,\dot{q}_d)$ (with $d > 1$) when 
\begin{equation}\label{trueSLfit}
\frac{\partial f}{\partial y} / \frac{\partial f}{\partial x} \pm \frac{dx}{dt} / \frac{dy}{dt}  \ \ \text{ is close to zero}
\end{equation}
for some pair $\{x,y\} \subseteq \{q_1,\ldots,q_d,\dot{q}_1,\ldots,\dot{q}_d\}$ of variables, and some choice of sign $\pm$.\footnote{Quoting end of \cite[Section S1]{SL2}: ``The partial derivative pairs define a cloud of
line segments in phase space, therefore we are only interested in matching the line but not
necessarily the direction of the line. Negating the $\Delta x/\Delta y$ term or taking the absolute value
of both can affect the signs of terms in the optimal law equation (for example, sign
differences between Lagrangian and Hamiltonian equations)."}  Given that their fitness function for two variables ($d=1$) satisfies this property (with the symbol ``$\frac{\delta f}{\delta y}$" interpreted as a partial derivative $\frac{\partial f}{\partial y}$), it is likely that (\ref{trueSLfit}) was the one used in their experiments.  It remains to understand how making the expression in (\ref{trueSLfit}) small could find laws in the system.  

Consider first the possibility that $y = \dot{x}$ and $x$ is a coordinate.  Then, expression (\ref{trueSLfit}) using the plus ``$+$" sign is zero when 
\begin{equation}\label{pluseq}
\dot{y} \frac{\partial f}{\partial y}  + y \frac{\partial f}{\partial x} = 0.
\end{equation}
Observe that if $f = \mathcal{H}$ is the Hamiltonian of the system, then $f$ satisfies Hamilton's equations: $\frac{\partial f}{\partial x} = -\dot{y}$, $\frac{\partial f}{\partial y} = y$.  In particular, $\mathcal{H}$ solves $(\ref{pluseq})$.  Notice also that for any differentiable function $g: \mathbb R \to \mathbb R$, we have that $f = g(\mathcal{H})$ 
also satisfies $(\ref{pluseq})$.  Thus, a high fitness should correspond to laws of the form $g(\mathcal{H})$; %(for instance constant multiples of $\mathcal{H}$); 
moreover, since $\frac{d g(\mathcal{H})}{dt} = g'(\mathcal{H})\frac{d \mathcal{H}}{dt} = 0$, these functions would be constants of motion.  Since the class of solutions to (\ref{pluseq}) is so large (it contains, for instance all power series in $\mathcal{H}$), it is unclear why it would hone in on a particular such function $f$, except that possibly the small-height operation tree setup of \cite{SL} constrains the complexity of $f$ so much that it is biased to find scalar multiples of $\mathcal{H}$.  

If instead, a minus ``$-$" sign had been chosen in (\ref{trueSLfit}), then one has an equation of the form $\dot{y} \frac{\partial f}{\partial y}  = y \frac{\partial f}{\partial x}$.
Suppose that $f = \mathcal{L} = T - V$ is the Lagrangian of the system with $T$ the kinetic energy and $V$ the potential energy.  If it turns out that the kinetic energy $T$ has the special form $\sum_{i=1}^d \frac{1}{2} m_i \dot{x}_i^2$, then $\frac{\partial f}{\partial y} = my$, and this fitness equation is equivalent to 
\begin{equation}\label{newton2ndlaw}
m \dot{y} = \frac{\partial V}{\partial x},
\end{equation}
which is Newton's second law for a force arising from a potential.  However, if $T$ does not have the special form indicated, then (\ref{trueSLfit}) will not find a Lagrangian.  As a simple example, the Lagrangian of a double pendulum system cannot be found in this way.  This is perhaps why the authors of \cite{SL} were unable to find the Lagrangian for a double pendulum system in their experiments (but were able to find its Hamiltonian).

In conclusion, once a sign is chosen in a modified fitness function (\ref{trueSLfit}), it is possible to find (non-canonical) functions of Hamiltonians.  When the opposite sign is chosen, it is possible in special circumstances to arrive at Lagrangians, but not possible in general.  Nonetheless, in each case, the natural law determined by the vanishing of this fitness measure is a consequence of classical physics embedded in the measure.  Given these considerations, it is unclear why one would not choose metrics specifically tailored to finding Hamiltonians and Lagrangians independently (such as those in Section \ref{HSfitness} below). 

\section{Fitness measures that find Hamiltonians and Lagrangians of a system}\label{HSfitness}

Based on theoretical considerations, we propose fitness criteria for finding Lagrangians and Hamiltonians of a physical system.  In classical physics, the Lagrangian $\mathcal{L}$ of a system with (generalized) coordinates $x_i$, $\dot{x}_i$ ($i = 1,\ldots,d$) solves the Euler-Lagrange equations:
\begin{equation*}\label{ELeqs}
 \frac{d}{dt} \left( \frac{\partial \mathcal{L}}{\partial \dot{x_i}} \right ) - \frac{\partial \mathcal{L}}{\partial x_i} = 0.
\end{equation*}
If we now assume that $\mathcal{L}$ is a function of the coordinates $x_j$ and their time derivatives $\dot{x_j}$, then by the chain rule, we have for each $i$ and all $t \in [0,a]$:
\begin{equation}\label{bigELeq}
EL_i(\mathcal{L},t) := \sum_{j=1}^d {\frac{\partial^2 \mathcal{L}}{\partial x_j \partial \dot{x}_i} } \frac{d x_j}{d t} +\sum_{j=1}^d {\frac{\partial^2 \mathcal{L}}{\partial \dot{x}_j \dot{x}_i}} \frac{d \dot{x}_j}{d t} - \frac{\partial \mathcal{L}}{\partial x_i} = 0.
\end{equation}
If we are given a function $f$ and discretized coordinate trajectory data coming from a physical system, then we may use $EL_i(f,t)$ as a measure of the \textit{Lagrangian fitness} LFit of a potential conservation function $f$.  That is, we compute symbolically the partial derivatives in (\ref{bigELeq}) and evaluate numerically the time derivatives $\frac{d x_j}{d t}$ and $\frac{d \dot{x}_j}{d t}$ over discretized time steps $t_k$ to calculate:
\begin{equation}\label{LagrangianFitness}
\text{LFit($f$)} := -\frac{1}{N} \sum_{k = 1}^N \log \left(1 + \sum_{i=1}^d \left | EL_{i}(f,t_k) \right| \right). 
\end{equation}

Consider now the equations for a Hamiltonian $\mathcal{H}$ of a physical system (which hold at all times $t$ of the trajectory and for each coordinate $i$):
\begin{equation*}\label{Heqs}
HQ_i(\mathcal{H},t) := \frac{\partial \mathcal{H}}{\partial \dot{x}_i} - \frac{d x_i}{d t} = 0, \ \ HP_i(\mathcal{H},t) := \frac{\partial  \mathcal{H}}{\partial x_i} + \frac{d \dot{x}_i}{d t} = 0.
\end{equation*}
Similar to above, we may measure the \textit{Hamiltonian fitness} HFit of a function $f$ over the discretized trajectory as follows:
\begin{equation}\label{HamiltonianFitness}
\text{HFit($f$)} := -\frac{1}{N} \sum_{k = 1}^N \log \left(1 + \sum_{i=1}^d \left | HQ_{i}(f,t_k) \right| + \sum_{i=1}^d  \left | HP_{i}(f,t_k) \right| \right). 
\end{equation}
It is straightforward to check that if HFit($f) = 0$ for all times $t$, then $f$ is a conserved quantity (as $f$ is then the canonical scale-dependent Hamiltonian of the system).
Clearly, however, the methods proposed here are limited to analyzing measurements given in canonical or generalized coordinates. 

% In implementations, one could also find laws by searching over subsets of the generalized coordinates of the given system (as described in \cite[Section S2]{SL2}).

\begin{figure}[htb]  % width=40mm,bb=0 0 800 800
  \makebox[\linewidth]{\includegraphics[scale=0.45]{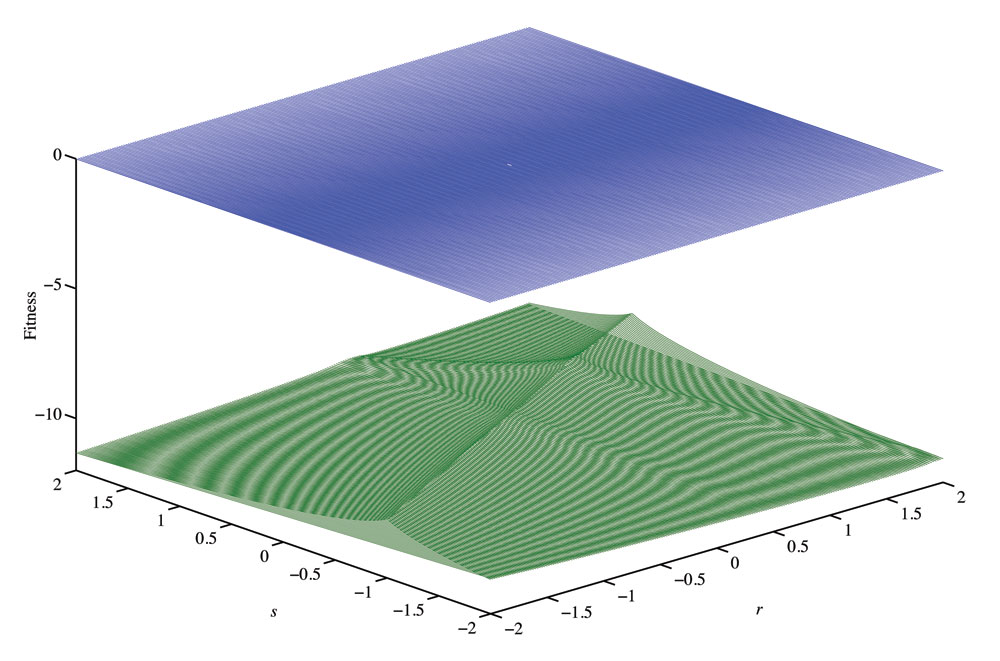}}
  \caption{
  Fitness functions for a harmonic oscillator.  The points in blue come from using (\ref{SLfitness}) while those in green are from (\ref{HamiltonianFitness}).}
\end{figure}\label{fig1}

For an experiment, we considered the simple harmonic oscillator system $\ddot{x} = -\omega^2 x$.  Its equation of motion is $x(t) = A \cos(\omega t + \phi)$, and its Hamiltonian is 
\begin{equation*}
\mathcal{H} = \frac{1}{2} \dot{x}^2 + \frac{1}{2} \omega^2 x^2.
\end{equation*}
Supposing candidate laws of the form $f =  \alpha \dot{x}^2+\beta x\dot{x}^2 + \gamma x^2$, we computed the fitness as a function of $\alpha$, $\beta$, and $\gamma$ using our metric (\ref{HamiltonianFitness}) and the one from  (\ref{SLfitness}). We used the motion-tracked data supplied online in the supplementary materials of \cite{SL2}.  

The optimal fitness using (\ref{HamiltonianFitness}) occured when $(\alpha,\beta,\gamma) = (.5,0,3)$.  In Figure \ref{fig1}, we plotted a linear section $\{\alpha = r-s $, $\beta = r-2s$, $\gamma = 5r/2+s\}$ of the different fitness measures as a function of two varying parameters $r$ and $s$ (the peak for the Hamiltonian fitness occurs at $r =1$, $s = .5$).  As predicted by (\ref{derivequality}), the fitness given by (\ref{SLfitness}) was flat.  When we used the altered fitness (\ref{trueSLfit}), however, we do achieve optimal fitness at the same parameters (not shown).

% ----------------------------------------------------------------

\end{document}